**Nonlinear price dynamics of S&P 100 stocks**

**Gunduz Caginalp**

**University of Pittsburgh**

**Pittsburgh, PA 15260**

**caginalp@pitt.edu**

**Mark DeSantis**

**Chapman University**

**Orange, CA 92866**

**desantis@chapman.edu**

**July 9, 2019**

**Abstract.** The methodology presented provides a quantitative way to characterize investor behavior and price dynamics within a particular asset class and time period. The methodology is applied to a data set consisting of over 250,000 data points of the S&P 100 stocks during 2004-2018. Using a two-way fixed-effects model, we uncover trader motivations including evidence of both under- and overreaction within a unified setting. A nonlinear relationship is found between return and trend suggesting a small, positive trend increases the return, while a larger one tends to decrease it. The shape parameters of the nonlinearity quantify trader motivation to buy into trends or wait for bargains. The methodology allows the testing of any behavioral finance bias or technical analysis concept.

<div align="center">**Nonlinear price dynamics of S&P 100 stocks**</div>

## 1. Introduction

Understanding the dynamics of asset prices holds the key to establishing or refuting many of the basic theories of economics and finance. Although there is no unique methodology for determining the valuation of a stock or asset, the basic principles are not in dispute. However, the changes in prices beyond those that can be explained by valuation have led to a lively debate. Classical finance provides for no other explanation than random fluctuations, while classical economics is largely a study of equilibrium conditions. Thus the study of the relative changes in asset prices, i.e. asset price dynamics, is a new discipline that can explain many of the ideas behind the motivations and strategies of traders and investors.

Key problems in finance related to asset pricing and dynamics can be broadly grouped into four categories arranged in order of increasing refinement and decreasing time scale:

I. Portfolio Optimization. Given a set of investments with a spectrum of variance and expected return, the investor strives to find an optimal portfolio of stocks, bonds and Treasury bills, etc. This is a classical problem in finance that usually assumes a time period of many years.

II. Risk Premium Analysis. An investor attempts to determine how the risk premium varies, i.e., ascertain when investors are more risk averse and move toward safer assets such as bonds or Treasury bills. This flight to or from quality generally involves a time period that varies from months to years.

III. Asset Price Dynamics. A trader tries to determine, for a particular asset, the likely change in price depending upon changes in valuation and the history of prices, volume, and volatility, among other factors. Such technical analysis is associated with a time scale of one day to several weeks (see also comments in Section 3.3.2).

IV.    Market Microstructure.  A trader attempts to determine the likely price change of a traded asset as a function of the order book.  The timescale is typically on the order of milliseconds to minutes for liquid securities and potentially hours for less liquid assets.

Most of classical finance focuses on (I) Portfolio Optimization, and thus, in providing a theoretical framework for choosing the highest return given a particular risk tolerance. However, a resolution to this problem provides little insight into the more refined issues further down this list. In recent years, there has been some effort to understand why investors as a group are more risk-averse during certain time periods than others, which is part of (II) Risk Premium Analysis. Empirically these preferences appear to vary on time periods of several months to many years and provide the key to understanding large market moves (e.g., the high tech bubble of the late 1990s).  This is an important line of investigation that is relevant to understanding significant aggregate market movements. However, it does not impact the more detailed issue of the factors behind price movements for a particular asset, namely problem (III), Asset Price Dynamics. Methodologies have been developed to understand how a stock should adjust to a new equilibrium price given changes in fundamentals, e.g., earnings forecasts.  However, the manner in which a stock should adjust to its recent price history, volume and volatility changes, resistance (upon nearing a recent high), etc. is largely untouched by the methodologies of (I) and (II).  Finally, the rapid price movements in actively traded stocks as orders flow into the marketplace have become another important issue (i.e., problem (IV) Market Microstructure) that has moved to the forefront as the debate rages on the effects of high-frequency trading.

Problems (I) - (IV) each illuminate different aspects of investor motivation. For example, (I) involves the key concept of risk tolerance in an equilibrium or static context, while (II) invokes temporal changes in risk tolerance together with other factors that lead to changes in the popularity of stocks. Problem (III) can illuminate investor and trader behavior and reactions on a daily basis that can provide valuable insight to the causes and evolution of critical periods of market instability, e.g., Fall 2008, as the housing crisis unfolded. Problem (IV) can be useful in developing market trading rules and their implications for order flow.

In this paper, we focus on problem (III). Behavioral finance has noted numerous diversions of asset prices from fundamental value and has provided evidence of the factors responsible for these discrepancies. For example, although both overreaction and underreaction have been established, advocates of classical finance argue that without a way to distinguish overreaction from underreaction, the ideas can be viewed as philosophical rather than practical. This study shows that both of these effects are present and provides clear criteria under which they can be expected. Hence, it is possible to distinguish between under- and overreaction so that it is clear when one or the other is more likely. It also demonstrates a new method that can illuminate and quantify the motivations of traders using market data. In particular, this methodology can examine either one stock or a class of stocks over a particular time period, and conclude, for example, whether momentum trading, profit taking, or changes in valuation motivates traders. Furthermore, these effects can be measured, and their relative importance examined. This methodology can be used for testing any particular motivation that is hypothesized and can be quantified.

It is widely acknowledged that "noise," or the randomness involved in valuation, complicates testing of hypotheses related to market dynamics (Black, 1986). Less well appreciated is the idea that examining nonlinear phenomena within the framework of linearity can obscure an effect completely. For example, suppose a market participant is examining the effect of the recent price trend on asset returns. If the empirical data is such that small uptrends have a positive effect on returns, while large uptrends have a negative effect, then a linear (unlike a nonlinear) regression may fail to establish a significant effect.

This paper pursues a methodology that overcomes both of these obstacles using the S&P 100 stocks. The first key idea is that by modeling the valuation and using this as one of the independent variables in a regression, much of the randomness that is inherent in valuation may be extracted. Second, the linear, square, and cubic terms of a suitably defined price trend variable are included as regressors. Thus, the effect of these terms on the return, which is the dependent variable, is obtained.

The primary objectives of this paper are to:

1. Provide a precise methodology for testing and quantifying the effect of any hypothesized influence beyond valuation on the dynamics of stocks by utilizing an equity valuation model together with non-classical effects such as trend and resistance.

2. Demonstrate that the short-term trend variable has a non-linear contribution to the daily return, i.e., small uptrends (downtrends) increase (decrease) daily return, while large uptrends (downtrends) decrease (increase) returns.

3. Demonstrate that this methodology can be used to examine long-standing practitioner strategies and trader behavior.

A natural goal for practitioners is the development of tools for forecasting. Our work provides a basis for this endeavor, but the goal in this paper is not to determine the best model for predictions but rather to examine quantitatively the existence of factors in price dynamics beyond valuation.

Closed-end funds (CEFs) and exchange-traded funds (ETFs) have been used in order to examine trader motivations (see Caginalp and DeSantis, 2011a, 2011b for closed-end funds; and Caginalp *et al.*, 2014 for ETFs). The key to this approach is the inclusion of an appropriately defined valuation variable as an independent regressor in the model. The net asset value (NAV) is utilized as a proxy for the fund's intrinsic value in these earlier studies. Since both CEFs and ETFs are special asset classes, it is important to determine whether investor behavior can be characterized for a broad class of major stocks that are of interest to both the individual and institutional investor.  In particular, we consider the stocks in the S&P 100.  As there is no unique and unambiguous valuation methodology for these stocks, an approach for estimating a stock's intrinsic value based upon market returns and forecasted earnings per share estimates is developed.  With this variable defined it is then possible to extend the methodology from prior works to stocks.  Thus, it opens the door to a new way of analyzing the price change dynamics of individual stocks and identifying additional influences on returns.  It is then possible to test any quantifiable variable in terms of its effect on price changes.

A discussion of the implications of the results in terms of the motivations in the marketplace and a description of how this methodology may be utilized to understand additional factors and strategies in the financial markets are presented in the Conclusion.

## 2. Background and Related Literature

### 2.1 Quantifying behavioral factors

Behavioral finance suggests a number of factors that influence price dynamics, but the establishment of these ideas in a quantitative and decisive manner is a new and evolving endeavor.  Among these behavioral factors are two opposing effects:  underreaction and overreaction.  A steady trend in prices is usually seen as evidence of underreaction whereby investors are slow to assimilate and act upon new information that arrives into the market stochastically.  Overreaction is essentially the opposite effect whereby traders bid up prices excessively as positive information enters the market, and analogously for negative information (Bremmer and Sweeney, 2001; Madura and Richie, 2004; Sturm, 2003).

Advocates of the efficient market hypothesis would argue that if both of these phenomena exist with similar frequencies in the marketplace, and there is no methodical way to distinguish them, then the market must be efficient (Fama, 1998).  Indeed, he suggests that any alternative model to the efficient market model must choose between under- and overreaction.  The analysis in Section 5, however, shows that both effects are present in this large data set of high capitalization, liquid stocks making up the S&P 100.  Furthermore, this paper provides a quantitative methodology to distinguish between them[1].  A cubic relationship exists between the return and the recent price trend, so that an uptrend has an increasingly positive effect on prices up to a maximum, at which point an increase in the uptrend diminishes the positive effect on return.  Beyond a particular point in the magnitude of the uptrend, there is a negative effect on the return.

Thus, the analysis demonstrates in a unified setting the impact of both underreaction and overreaction within a data set of large capitalization stocks. The distinction between these is determined empirically. Among the other variables considered in the analysis is volatility (both short- and long-term), which is a key concept in finance that is generally assumed to be a proxy for risk.  Both volatility variables have statistically significant and positive regression coefficients.  In addition, daily trading volume exhibits a positive and marginally significant impact on the return.

This paper utilizes a data set comprising daily closing prices of the S&P 100 stocks from 2004-2018.  While many studies of stock prices have focused on yearly or monthly data, there are a number of advantages to studying the shorter time periods from the perspective of understanding the motivations of traders and investors.  One of these is that the short time scale reduces the impact of long term issues such as economic cycles, demographics and other influences on prices.  Another involves the size of the data.  Since many stocks are highly correlated with one another, using a time scale of years severely reduces the amount of data.  Examining this issue further, we consider the time period 1998 to 2018.  Most stocks experienced a sharp rise during the first two years, then a sharp decline for three years;

---

[1] Bloomfield *et al.* (2000) attempt to discern between situations in which one might expect under- or overreaction by conducting laboratory market experiments.  Their results suggest this can be accomplished provided one knows the "reliability of investors' information."  Indeed, as the reliability of information increases both "prices and investors' value estimates tend to underreact."

followed by another boom during 2003-2007 and another sharp decline during 2007-2009; followed by a strong recovery during 2009-2018. Even if one considered the yearly dynamics of 1,000 stocks during this 20 year period, one would essentially be observing these three cycles and arriving at conclusions that describe this era, but not necessarily the fundamental forces in stock dynamics. Thus, the use of yearly data is equivalent to studying a small data sample.

*2.2 Asset pricing and estimation of intrinsic value*

Several studies, for example the surveys of Subrahmanyam (2010) and Goyal (2012), note that a tremendous amount of research has been devoted to understanding why one stock might exhibit a higher expected return than another. At the heart of these cross-sectional studies lies the Fama-Macbeth (1973) regression methodology and the development of n-factor models.

Our goal, however, is to determine the variables beyond valuation that impact *future* returns on a daily basis and to develop a methodology that facilitates understanding of trader motivations and strategies.

A first step in this process is the estimation of the value of a stock. This a classical problem in finance, dating back to Graham and Dodd (1934) [see also review by Pinto et. al. (2015)]. In principle, one would like to estimate the daily valuation as with the closed-end and exchange-traded fund studies in which one can approximate the intrinsic value through the use of a proxy such as the fund's net asset value. Lee *et al.* (1999) notes that "the scant attention paid to this important topic [measurement of the intrinsic value] reflects the standard academic view that a security's price is the best available estimate of the intrinsic value."

One way to account for much of the daily change in valuation is by assuming that the relative change in a stock's price should, on average, be proportional to the relative change in the overall market.  This concept of beta has long been used by practitioners and academics[2].  Moreover, cross-sectional regression studies of asset prices consistently control for the return on the market.

A study by Ang and Bekaert (2007) finds that the earnings yield is a strong predictor of future cash flows, but does not help to explain excess returns.  Bali *et al.* (2008) find that "earnings yield is a significant predictor of firm-level stock returns..."  Campbell and Thompson (2008) find that several valuation ratios, including earnings-to-price, serve as better out-of-sample predictors for returns than historical return data.  Lee *et al.* (1999) find that the use of consensus analyst earnings forecasts yield better predictive results than using time series methods on historical data.  Thus, rather than using the current earnings yield which does not contain new or forward-looking information, the relative change in the expected earnings per share is utilized.  Although Ang and Bekaert (2007) find that the short-term interest rate is a "robust predictive variable for future excess returns …at short horizons", we utilize a long-term rate.  While the Ang and Bekaert (2007) data set includes data up to 2001, our data set consists of data from December 2004 to July 2018.  Note that for much of our time period of interest short-term interest rates were held artificially low (near zero).  Thus, we believe that relative changes in the longer term rate are more informative as a 5% change in the short-term rate (e.g., 0.21% versus 0.20% are essentially the same for an equity investor) is not as significant as a 5% change in the long-term rate[3] which was more significant during this time period. While long-term rates were also targeted by the Federal Reserve during this time period, these rates were generally higher and provided some competition for stocks. Finally, Vassalou (2003) shows that news regarding upcoming gross domestic product (GDP) growth can help to explain the cross-section of returns. So relative changes in the forecasted GDP are utilized (see below).

---

[2] In addition to the numerous academic articles regarding beta, standard Finance textbooks such as Bodie et al. (2008) and Luenberger (1998) also cover this topic.  From a practitioner's perspective, beta's importance is evident from its inclusion on the standard equity summary screens/pages of, for example, Bloomberg and Value Line.

[3] Further support for the long-term bond yield is provided by Neely, et al. (2014) who find that it helps to predict the equity risk premium.

In order to quantify the changes in valuation[4], we consider a set of variables that practitioners have long known to be relevant and recent studies have established as statistically significant.  There are numerous effects on stocks. These can be divided generally into two sets: those that impact the overall market and those that are firm specific. Since the former are already incorporated into the price changes in the S&P 500, we can utilize these changes rather than consider each of the numerous indicators that impact the market.  The change in the valuation of a particular stock can be separated into three parts:

(i)     Changes due to adjustments in the overall economy.  Any modifications to the GDP forecast will have an impact on expected output, and hence profits for the aggregate market. Changes in interest rates affect borrowing costs and consumer purchases.  Each company will feel the influence of these two key indicators, and the magnitude of the impact will depend on the type of industry.  When determining each stock's valuation, the regressions include a coefficient for each of these two variables to measure that magnitude.

(ii)    Changes to the aggregate market due to exogenous events. There are many other factors that influence the stock market beyond GDP, interest rates, and similar variables.  These include political events, extreme weather, earthquakes, tsunamis, etc. that tend to push the entire market up or down.  By using the S&P 500 as a proxy for the aggregate market and including a coefficient for the change in the S&P 500, we can compensate for this effect, as well as the standard effects described in (i).  We use the S&P 500 rather than the S&P 100 since it is a broader measure of market activity.

(iii)   Changes due to firm-specific events.  Company-specific news will impact a particular stock beyond that of the overall market.  While there are many events that can change the perceived value of a company, the forecast of future earnings is the most important, as this essentially determines the value of the company by classical measures.  Analysts are constantly updating their forecasts for the companies they cover, and for each stock there is (on average) a couple of dozen analysts whose average forecast can easily be found in financial news services on the internet[5].  By including a coefficient in the regression for the earnings, we capture this key factor for the firm-specific valuation.

---

[4] A detailed description of the valuation methodology is included in Section 3.2.1.

As such, the effects of the overall market, interest rates, and both *forecasted* earnings per share as well as *forecasted* GDP are incorporated into the valuation variable[6].

## 3. Data Set and Variable Definitions

### 3.1 Data Set

The data set is a balanced panel of 279,565 daily observations corresponding to the date range December 28, 2004 through January 18, 2018.  The observations are for 85 of the S&P 100 companies (as of January 20, 2018) with 3,289 records per stock, i.e., all stock for which complete data was available in this set. Given the definitions of certain variables[7], and the absence of data in some cases, the regressions are run on a balanced panel of 257,635 observations (3,031 records per stock covering the time period January 3, 2006 through January 17, 2018).  These are large, highly liquid firms.  Indeed, the median firm has an average daily market capitalization of $96,039,310,000 and an average daily

---

[5] While these forecasts do not change on a daily basis, approximately 27% of the daily observations in our data set contain revisions of earnings' forecasts.  The fact that there is no change on the remaining 73% of the days does not diminish the importance of this information in the regressions.

[6] There exists a long history of practitioners' use of these variables in addition to the academic support noted in Section 2.2.  If any of these variables are not significant factors in the stock's valuation, then the regressions will produce zero coefficients.

[7] Due to insufficient data, we exclude CHTR from all regression analyses.  GOOG and FOXA are excluded, while GOOGL and FOX are included.  The resulting unbalanced data set consists of 283,153 daily observations corresponding to 98 companies.  To produce our primary analyses with a balanced data set, we also exclude the following two groupings of stocks.  First, as the Long Term Trend and Volatility variables require a year of data to calculate, we excluded the following stocks due to a lack of pricing data: ABBV, AMZN, CELG, FB, GM, KHC, KMI, MA, PCLN, PM, PYPL, SPG, and V (earliest records are post December 28, 2004).  Second, we excluded DWDP due to a lack of forecasted earnings per share data.  Refer to Table 3 for company names.

dollar trading volume of $577,130,000 (see Table 1). Hence, this set comprises an unbiased collection of large capitalization and liquid stocks. The set is large enough to obtain broad results while still being within the computing capacity. Note that since we are examining daily (rather than yearly) changes, it is unlikely that neglecting those companies for which adequate records are available would introduce survivorship issues. This would likely be a more relevant issue if we were examining mutual fund returns over years. In that case, one might have, for example, underperformance in the funds that have been terminated.

<< Insert Table 1 here >>

*3.2 Variable Definitions*

This paper's methodology, which is described in Section 4, consists of running two-way fixed effects regressions with the following day's return as the dependent variable. That is, if today is day *t*, then the dependent variable is defined as

$$R_{i,t+1} = \frac{P_{i,t+1} - P_{i,t}}{P_{i,t}}$$

where $P_{i,t}$ is the daily adjusted close price for firm $i$. This price, which accounts for dividends, is obtained from Bloomberg's "Total Return Index Gross Dividends" field. The regressions' independent variables are based upon this closing price and are described in the remainder of this section.

*3.2.1 Valuation*

The determination of the Valuation variable is a two-step process. Given a stock $i$, for each day $t$ an ordinary linear regression is performed for the time period $[t - 188, t]$ using the relative changes in the (Bloomberg) forecasted earnings per share for stock $i$ for the current year, $EPS_{i,t}$; the return of the S&P 500 Index (SPX), $MKT_t$; the yield on 10-year U.S. Treasury bonds, $INT_t$; and the forecasted gross domestic product for the current[8] year, $GDP_t$, as independent variables[9]. Note that we are using the relative changes in these quantities since it is the change in the expectation on which traders focus. The earnings per share variables reflect the forecasted earnings per share data as of day $t$. Thus, earnings updates/revisions are reflected in this variable. The $MKT_t$ variable is a proxy for the market return. The stock's return, $R_{i,t}$, for stock $i$ on day $t$ is utilized as the dependent variable. The regression is thus of the form

$$R_{i,t} \sim \alpha_i^{(0)} + \alpha_i^{(1)} EPS_{i,t} + \alpha_i^{(2)} MKT_t + \alpha_i^{(3)} INT_t + \alpha_i^{(4)} GDP_t + \theta_{i,t} \tag{1}$$

where $\theta_{i,t}$ is an error term and the parameters $\alpha_{i,t}^{(j)}$ correspond to stock $i$ on day $t$ for independent variable $j$. Further analysis of the error terms, which are close to Gaussian between the 1% and 99% levels, can be found in the Appendix. The nine-month period was chosen as on optimal time period as a short time period does not provide enough data for a reliable set of regression coefficients. Examining data that is much older can provide data from a different era. For example, in 2009, the data from the crisis may be less relevant than the post-crisis data.

These coefficients for this rolling nine-month period are used to determine a valuation for time $t$ using the linear equation:

$$Val_{i,t} = \left( \hat{\alpha}_{i,t}^{(0)} + \hat{\alpha}_{i,t}^{(1)} EPS_{i,t} + \hat{\alpha}_{i,t}^{(2)} MKT_t + \hat{\alpha}_{i,t}^{(3)} INT_t + \hat{\alpha}_{i,t}^{(4)} GDP_t + 1 \right) * P_{i,t-1} \tag{2}$$

---

[8] Although the GDP estimates are less volatile toward the end of the year, changes in forecasts for the last quarter continue to dominate expectations relating to growth or slowdown. The role of GDP changes is minor in any case as the outlook for S&P earnings is quickly incorporated into the MKT variable.

[9] The yield on the 10-year U.S. Treasury bill was taken from Bloomberg (U.S. Generic Government 10 Year Yield). Forecasted GDP data for the current calendar year was obtained from *The Economist* magazine.

where $\hat{a}_{i,t}^{(j)}$ represents the estimated regression coefficient from equation (1). Thus, equation (2) provides an estimate of the value of stock $i$ on day $t$ based on its relationship with the market index and its earnings outlook.[10] Note that the information used at time $t$ involves only the days prior.

The Valuation variable, which is the relative excess value (and measures the extent of undervaluation) is defined via the expression

$$Valuation_{i,t} = \frac{Val_{i,t} - P_{i,t}}{Val_{i,t}}$$

This affords the opportunity to test the hypothesis that the daily return is greater when this independent variable is positive. This is consistent with the earlier regression studies cited above.

The nature of the marketplace is that traders are always trying to obtain the most recent information about the present and future in order to gain an edge in making profits. For example, the earnings per share for the previous year is already incorporated into the market, and, thus is not as important as the earnings report that will be issued in the future. Thus, traders are focused on the prospects for future earnings and any signs of changes. In utilizing information about the earnings for each company, the change in the prevailing forecasts on the day of the trade is considered. It would appear that taking a weighted average of the forecasts for the current year and the following year could yield a more precise estimate. In other words, on day $d$ of the 252 trading days, the weight given to the current year would be $(252 - d)/252$, with the remainder of the weighting assigned to the following year. We re-ran regression models 1V, 2, 3, and 4 with the valuation variable calculated with weighted averages for EPS and GDP. The results are qualitatively similar to those reported in Table 2. This is not surprising as the two valuation variables are highly correlated (with the correlation computed on a per firm basis). Indeed, the median correlation is 0.99 and the minimum correlation is 0.94.

---

[10] If a coefficient is not statistically significant (p-value >= 0.10), then it is set to zero in equation (2).

*3.2.2 Price Trend*

Several prior empirical studies have noted the existence of a trend in asset prices.  See, for example, Carhart (1997), Jegadeesh and Titman (1993), Moskowitz and Grinblatt (1999), and George and Hwang (2004).  In addition, several theoretical models, starting with Caginalp and Ermentrout (1990) using differential equations have been developed to model price behavior when traders are motivated by both trend and valuation. In addition, Barberis, Shleifer, and Vishny (1993), Daniel, Hirshleifer, and Subrahmanyam (1998), and Hong and Stein (1999), have developed models to relate intermediate-term trend coupled with long term reversals.

Moreover, studies have also identified phenomena such as underreaction and overreaction in asset prices. Bremer and Sweeney (1991), Madura and Richie (2004), Sturm (2003) define underreaction (overreaction) as positive (negative) returns following large positive price movements and negative (positive) returns following large negative price movements.

Practitioners also recognize the impact of the recent trend in price on asset prices through sayings such as "The trend is your friend."  Indeed, a survey by Menkhoff (2010) suggests that practitioners frequently use technical analysis (including, in particular, trend following) for trading horizons of weeks. While the efficient market hypothesis and technical analysis seem to be in conflict, Sturm (2013) concludes that the latter "…attempts to measure changes in these beliefs to predict stock prices and should have value given the evidence in behavioral finance…"  In other words, the process of price discovery as new information enters the market provides an opportunity for technical analysis to be useful.  Moreover, traders may be inclined to act on deviations from the recent trend rather than the trend itself[11].  For example, suppose returns have been trending upward at 0.5% per day over the past two weeks.  If today's relative change in price is consistent with that trend, then traders may not respond. However, if today's relative change in price surges to, say, 1%, then traders may act. In this case traders are responding to the "acceleration" in returns rather than the "derivative" of returns.

---

[11] Various technical analysis methods focus on the differences of moving averages that, in effect, are a measure of a smoothed out second derivative, i.e., the change in the trend over a particular time period.

Given these two factors, i.e. that stock prices tend to exhibit a trend and traders may respond to deviations from the recent trend more so than the trend itself, the trend variable is defined[12] as follows:

$$Trend_{i,t} = \frac{P_{i,t} - P_{i,t-1}}{P_{i,t-1}} - \frac{1}{0.58195} \sum_{k=1}^{10} e^{-k} \frac{P_{i,t-k} - P_{i,t-k-1}}{P_{i,t-k-1}}.$$

This Price Trend variable for firm $i$ is defined by subtracting an exponentially weighted average of the past ten days' relative changes in price from the current day's relative price change. This modeling approach is similar to the manner in which the valuation variable is defined in prior studies[13].

A key aspect of this paper's analysis involves the nonlinear relationship between the recent trend and the following day's return. This is achieved by including the square and cube of this trend variable in the regressions described in Section 4.

### 3.2.3 Volatility

In classical finance, risk is often associated with the standard deviation of returns. In order to understand the impact of volatility without the trend effect, the Volatility variable for firm $i$ is defined as the standard deviation of returns over the past $X + 1$ days, i.e.,

$$Volatility_{i,t} = \left\{ \frac{1}{X} \sum_{k=t}^{t-X} \left[ R_{i,k} - Mean(R_{i,[t-X,t]}) \right]^2 \right\}^{1/2}$$

---

[12] The fraction, $1/0.58195$, is a normalization factor equal to the inverse of the sum $\sum_{k=1}^{10} e^{-k}$.
[13] The price trend definition is reminiscent of the moving averages utilized in prior works (see, for example, Neely, et al., 2014), though these moving average definitions typically assign equal weight to each observation.

where $X = 10$ for short-term volatility and $X = 251$ for long-term volatility. Previous studies on closed-end (Caginalp and DeSantis, 2011a, 2011b) and exchange-traded (Caginalp *et al.*, 2014) funds have shown an ambiguous role for Volatility. With respect to closed-end funds Long Term Volatility tended to depress returns, while an increase in Short-Term Volatility boosted closed-end fund returns. For ETFs, both long- and short-term volatility increased returns (though the longer term volatility was only marginally significant). These variables are included in the current study to examine their effect on stock returns.

### 3.2.4 Long Term Trend

Several studies (e.g., Poterba and Summers, 1988) have found that returns tend to regress to the mean or experience price reversals over longer time periods. Theoretical justification has been shown in Leung and Wang 2018. We test for the existence of this phenomenon. Indeed, to test this hypothesis an annual trend variable is defined as the slope of the straight line fitted to returns over the past year scaled to annual units. That is, the variable, $Long\ Term\ Trend_{i,t}$, is the slope of the straight line fitted to $(P_{i,t} - P_{i,t-1})/P_{i,t-1}$ multiplied by 251, the number of trading days per year. Hence this is an average of the price changed during the previous 251 trading days.

### 3.2.5 Resistance

Consider the scenario in which individuals experience regret upon failing to capitalize by selling at the recent high price. These individuals might be inclined to sell if the price recovers to again approach this recent value that has become "anchored" in their minds (in analogy with Tversky and Kahneman, 1974). This selling exerts a downward pressure on the price. If this downward force keeps the price from attaining its recent high, then the price has encountered resistance. That is, the notion that an increase in price tends to slow or even retreat as the price approaches a recent high price is termed Resistance. Studies by George and Hwang (2004) and Sturm (2008) found evidence of Resistance in stock prices on monthly and annual time scales by forming portfolios based on the difference between the current stock price and a recent high price.

The binary Resistance variable is defined as follows:

(i)      Let $H_{i,t} \coloneqq max[P_{i,s}] \ for \ s \ in \ [t-63, t-16]$ and

(ii)      Set the Resistance variable if the following hold

         (1)   $for \ s \ in \ [t-15, t-10], P_{i,s} \leq 0.85 H_{i,t}$ and

         (2)   $0.85 H_{i,t} \leq P_{i,t} \leq H_{i,t}$.

That is, Resistance is set for day $t$ if the recent price has been less than 85% of the recent quarterly high price and the current price is between 85% of this recent high price and the high price. Approximately 1% of the records in the data set (2,139 out of 257,635) satisfy these criteria.

*3.2.6 Volume*

Market participants have long recognized the significance of trading volume in financial markets. As noted in Karpoff (1987), "It takes volume to make prices move." Numerous theoretical (e.g., Blume, Easley, and O'Hara, 1994; Harris and Raviv, 1993; Wang 1994) papers have developed models in which volume and the magnitude of price changes are positively linked. Karpoff (1987) provides a survey of papers and provides empirical evidence that volume is positively related to the magnitude of price changes and to the price change itself (the latter holding only in equity markets). Antoniou *et al.*, (1997) finds evidence suggesting that volume might help to predict returns.

Given the extensive prior literature and the practitioners' viewpoint, the recent trend in daily dollar trading volume for firm $i$

$$Volume_{i,t} = \frac{1}{0.58195} \sum_{k=1}^{10} \frac{Turnover_{i,t-k+1} - Turnover_{i,t-k}}{Turnover_{i,t-k}} e^{-k}$$

is utilized as the volume variable. $Turnover_{i,t}$ corresponds to Bloomberg's dollar trading volume for firm $i$ on day $t$. This variable is defined as the sum over all transactions for day $t$ of the product of shares traded multiplied by the trading price for each transaction. Dollar trading volume has been utilized in prior studies as a proxy for liquidity (see, for example, Chordia *et al.*, 2001).

## 4. Methodology

The data set consists of 85 time series - one for each company in the study. These time series are appended to one another and grouped by stock. Thus, the data set may be characterized as a time-series-cross-sectional data set or a panel data set (though, it has a large number of both cross-sections, 85 firms, and time periods, 3,031 days). Asset pricing models are typically tested with time series or cross-sectional methods (Goyal, 2012). For example, the Fama-Macbeth regressions account for contemporaneous correlations but not serial correlation (over time) in the error terms. Bali *et al.* (2008) utilize a fixed-effect regression with the firm as the sole fixed effect to consider the predictive relationship between earnings and stock returns at the firm-level.

Both firm heterogeneity and contemporaneous correlations are accounted for by utilizing a two-way fixed effects model with both firm and time fixed effects[14]. An advantage of the fixed-effects model is that it mitigates the impact of omitted variable bias. In addition, the unobserved firm-specific effects may be correlated with the independent variables. Although use of this model restricts any inferences to the subset of stocks considered (S&P 100) in contrast to a random-effects model, it should be noted that as the number of time periods per firm increases, the random effects estimator approaches the fixed effects estimator (Hsiao, 1996). As the data set contains 3,031 time observations per firm, these estimators should be similar. The applicability of this method, particularly for contemporaneous correlations, is described in Caginalp *et al.* (2014).

---

[14] This is accomplished by utilizing the FIXTWO option of the SAS PANEL procedure.

Moreover, as noted in Section 2, most asset pricing studies account for the three Fama-French factors (and frequently the momentum factor as well). As these factors vary over time but not across firms, the model is not enhanced by their inclusion. However, use of the time fixed effect controls for these (and any other firm invariant) factors. Thus, this paper's results are robust to any time and/or firm invariant factors.

The primary goal of this study is to analyze and better understand the factors that underlie the dynamics of stock prices (i.e., the daily changes in price). This paper considers the effect of the variables discussed in Section 3 defined at time $t$ on the following day's return, $R_{i,t+1}$. To that end, we consider regressions of the form

$$R_{i,t+1} = \alpha_0 + \alpha_1 Valuation_{i,t} + \alpha_2 Trend_{i,t} + \cdots + u_{i,t}$$

where

$$u_{i,t} = \mu_i + \gamma_t + \varepsilon_{i,t}$$

with $\mu_i$ representing the firm-specific effect, $\gamma_t$ the time-specific effect, and $\varepsilon_{i,t}$ the idiosyncratic error term.[15]

To minimize the impact of outliers in the data, all independent variables (with the exception of the Resistance variable) are winsorized at the 99th percentile[16]. As stock prices commonly exhibit "fat tails," this helps to minimize the impact of a few extreme observations. Without winsorizing the data, it would be possible for a few values to distort the motivations that dominate trading most of the time. A future topic of study would be to focus on these extreme events that often occur in the midst of a crisis.

---

[15] We report the R-square measure of Theil (1961) for all regressions. The R-square values are fairly large, which is not surprising given the fixed effects for firm and time. However, what is most important to our analyses is the significance of the coefficients of the primary explanatory variables, e.g. Valuation, Price Trend, etc., rather than goodness of fit of the model.

[16] That is, all outliers above (below) the 99th (1st) percentile are set to the value corresponding to the 99th (1st) percentile.

For the most comprehensive model, the results without winsorizing (reported in Table 4 in the Appendix) retain the vast majority of the coefficients and the basic features (i.e., sign and statistical significance) with different coefficients, of course.

Values for different variables may differ by orders of magnitude. Consequently, after winsorizing the data, each independent variable (with the exception of the indicator Resistance variable[17]) is standardized by firm. This is accomplished by first subtracting the firm's mean value and then dividing by the standard deviation. This facilitates comparisons of regression coefficients, which reflect a variable's effect on the following day's return, by utilizing the standard deviation as a natural scale of measurement. For example, consideration of the coefficients for Long Term Volatility (0.261) and Long Term Trend (-0.290) from Model 4 in Table 2 shows that a one standard deviation increase in the Long Term Trend is large enough to counteract the effect of a one standard deviation increase in the Long Term Volatility variable on the following day's return.

We utilize robust standard errors to account for heteroscedasticity[18]. We also cluster the standard errors to account for the fact that the regression residuals might be correlated by firm and/or time (Petersen, 2009; Thompson, 2011). Thompson (2011) notes that if the firm and time dimensions are not equivalent, then it is preferable to cluster along the dimension with fewest observations. As each of the 85 firms in this study has 3,031 daily observations, clustering is performed along the firm dimension to account for within-firm correlations.

---

[17] Resistance is a highly skewed binary variable. That is, only 2,139 out of 257,635 observations meet the Resistance criteria. Thus, standardization by a single standard deviation may not be sufficient to obtain regression coefficients comparable to those of the other variables (Gelman, 2008). Therefore this variable is left unstandardized. As such, note that the regression coefficient of this variable should be interpreted with care when comparing to the coefficients of other variables.

[18] There exist several heteroscedasticity-consistent covariance matrix estimators. Davidson and MacKinnon (1993) describe four such alternatives and infer that option 3 may perform best. As such, we employ this estimator which utilizes the squared residuals divided by $(1 + h_t)^2$ to estimate the diagonal entries of the error covariance matrix. The variable, $h_t$, is defined as $X_t(X^T X)^{-1}X_t^T$ where $X$ is the regressor matrix with $t^{th}$ row denoted by $X_t$.

**5. Regression Results**

To ensure our valuation variable is appropriately defined, we first regress the dependent variable, return on day $t + 1$, against the change in valuation on day $t$, using the definition of $R_{i,t+1}$ in Section 3.2. Intuitively, we expect the coefficient of the valuation variable to be positive. To confirm this we consider Model 1V

$$R_{i,t+1} = \alpha_0 + \alpha_1 Valuation_{i,t} + u_{i,t}$$

where $u_{i,t}$ is as defined in Section 4. The results of this model are reported in Table 2. Consistent with our expectation the coefficient of the valuation variable is positive.

Since our primary interest is the nature of the relationship between the return and the recent trend in price, we next consider Model 1T

$$R_{i,t+1} = \alpha_0 + \alpha_1 Trend_{i,t} + u_{i,t}.$$

The coefficient of the sole independent variable, *Trend*, is positive as one might intuitively expect, and statistically significant.

<< Insert Table 2 here >>

As described in Section 2.2, it is hypothesized that the inclusion of an appropriately defined valuation variable will account for much of the "noise" in returns and allow the effect of other variables, for example the recent trend, to be noted and measured. To this end, we combine Models 1V and 1T and consider Model 2

$$R_{i,t+1} = \alpha_0 + \alpha_1 Valuation_{i,t} + \alpha_2 Trend_{i,t} + u_{i,t}.$$

The results of Model 2 are given in Table 2. The coefficient of the valuation variable is both statistically significant and positive. Consistent with the results from Model 1T, the Price Trend's coefficient is still positive and statistically significant. These results also suggest that a small increase (decrease) in the price trend corresponds, on average, to a small increase (decrease) in the following day's return. However, it remains to be determined if this effect is truly linear[19].

We next consider the nature of the relationship between the following day's return and the valuation and trend variables. As previously noted, using a purely linear regression would likely mask important effects that can only be observed with nonlinear terms. Indeed, if a straight line is fit to the curve $y = x^3$, then a coefficient of zero would be obtained for the linear term. While this coefficient may be significant, the linear approximation is certainly inappropriate and misleading. Thus, we seek to determine whether the relationship between trend and return is nonlinear, similar to those of closed-end and exchanged-traded funds. To this end Model 3 builds upon Model 2 by incorporating the square and cube of the trend and valuation variables along with their higher order cross terms (up to third order). Model 3 has the form

$$R_{i,t+1} = \alpha_0 + \alpha_1 Valuation_{i,t} + \alpha_2 Valuation_{i,t}^2 + \alpha_3 Valuation_{i,t}^3 + \alpha_4 Trend_{i,t} + \alpha_5 Trend_{i,t}^2$$
$$+ \alpha_6 Trend_{i,t}^3 + \alpha_7 Trend_{i,t} * Valuation_{i,t} + \alpha_8 Trend_{i,t}^2 * Valuation_{i,t}$$
$$+ \alpha_9 Trend_{i,t} * Valuation_{i,t}^2 + u_{i,t}.$$

Using the quadratic and cubic terms provides enough degrees of freedom to describe the general shape of the dependent variable as a function of the independent. This is analogous to using the first three terms of a Taylor series for a function.

The results of Model 3, which are included in Table 2, indicate that the coefficients of the Valuation and Trend variables remain significant and positive. The coefficient of the quadratic Valuation term is also significant and positive suggesting that large changes in the daily trend correspond to positive returns the following day. Although the coefficient of the quadratic Trend variable is not significant, the coefficient of the cubic Trend variable is both significant and negative. This suggests that large positive (negative) changes in the daily trend tend to lead to lower (higher) returns. Indeed, consider Figure 1, which displays the return as a function of both the Valuation and the Trend. That is, using the statistically significant coefficients from Model 3, the following function

$$R_{t+1}(Valuation_t, Trend_t)$$
$$= 0.615 Valuation_t + 0.112 Valuation_t^2 + 0.721 Trend_t - 0.090 Trend_t^3$$
$$+ 0.151 Trend_t * Valuation_t^2.$$

is plotted.

---

[19] Augmenting Model 2 with an interaction term (Valuation*Price Trend) has a negligible effect on the significance, magnitude, and sign of the coefficients for the Valuation and Price Trend variables. Indeed, the Valuation coefficient is 0.000399 with a p-Value of 0.0004, while the Price Trend coefficient is 0.000577 with a p-Value less than 0.0001. The interaction term's coefficient, -0.00009, is not significant with a p-value of 0.2572.

<< Insert Figure 1 here >>

Restricting focus to the scenario in which the change in Trend is assumed to be zero[20], the cross-section of Figure 1 displayed in Figure 2 is obtained. This curve represents the following day's return, $R_{t+1}$, as a function of the Valuation variable. This curve is fairly linear for changes in Valuation greater than negative one standard deviation and somewhat flat for smaller changes. Consistent with intuitive expectations, returns tend to increase with increasing valuation. Though, larger negative changes in valuation tend to correspond to similar changes in return[21].

<< Insert Figure 2 here >>

---

[20] As the trend and valuation variables are standardized, considering a trend or valuation of zero is equivalent to considering the variable assumes its average value. Figures 2 and 3 are included as representative graphical depictions of the relationship between the return and valuation or trend.

[21] For clarity all figures are plotted without an intercept as this term would not change the shape of the curve (surface), but would merely shift it vertically. Thus, the focus is on local maximum and minimum values as well as the return at different values of the independent variables. For example, in Figure 3 a return, $R_{t+1}$, of 0.25 is achieved at Price Trend values of -2.99, 0.352, and 2.64, respectively.

As noted above, both academics and practitioners concede the recent trend in price plays a key role in asset returns. Consider the Price Trend terms in Model 3 of Table 2. Note the coefficient of the linear Price Trend term is statistically significant and positive, consistent with the intuition that rising prices tend to continue to rise. The cubic trend term's negative coefficient suggests that larger (smaller) price trends lead to smaller (larger) returns. Indeed, taking a cross-section of Figure 1 with the Valuation variable set to zero yields the cubic curve in Figure 3, which depicts $R_{t+1}$ as a function of the Trend variable.

This curve has a local minimum of -0.785 at a trend value of -1.634 standard deviations and attains a local maximum of 0.785 at a trend value of 1.634. Thus, as trend values increase from -1.634 to 1.634 standard deviations, the following day's return also increases. However, once the trend increases [decreases] beyond 1.634 [-1.634] standard deviations, then the return decreases [increases]. Thus, the data exhibit evidence supporting both underreaction and overreaction to price changes. Here underreaction is defined as a continuation of the trend and overreaction as a reversal of the trend.

<< Insert Figure 3 here >>

Earlier studies on different time periods revealed an asymmetry in the local minimum and maximum, suggesting that traders were more eager to take profits than buy on dips (Caginalp *et al.*, 2014). However, Figure 3 is quite symmetric in this regard. The symmetry may be a consequence of the long time period of this study that includes both bull and bear markets. The level of symmetry may be an indication of general investor caution or optimism during a particular period.

As noted in Section 3, numerous variables have been hypothesized to impact returns. Model 4 augments Model 3 with a subset of these factors. Model 4 has the form

$$R_{i,t+1} = \alpha_0 + \alpha_1 Valuation_{i,t} + \alpha_2 Valuation_{i,t}^2 + \alpha_3 Valuation_{i,t}^3 + \alpha_4 Trend_{i,t} + \alpha_5 Trend_{i,t}^2$$

$$+ \alpha_6 Trend_{i,t}^3 + \alpha_7 Trend_{i,t} * Valuation_{i,t} + \alpha_8 Trend_{i,t}^2 * Valuation_{i,t}$$

$$+ \alpha_9 Trend_{i,t} * Valuation_{i,t}^2 + \alpha_{10} Short\ Term\ Volatility_{i,t}$$

$$+ \alpha_{11} Long\ Term\ Volatility_{i,t} + \alpha_{12} Long\ Term\ Trend_{i,t} + \alpha_{13} Volume_{i,t}$$

$$+ \alpha_{14} Resistance_{i,t} + u_{i,t}.$$

Whether volatility in returns, longer term price trend, dollar trading volume, and proximity to a recent high price have an effect on the return is tested. While the Long Term Volatility variable exhibits a statistically significant, positive coefficient, the Short Term Volatility variable's coefficient is not significant. This is not consistent with the results on exchange traded funds in Caginalp *et al.* (2014) in which the Short Term Volatility variable is significant with a positive coefficient, and the Long Term Volatility is not significant. The discrepancy may reflect the greater use of ETFs for rapid trading and re-adjustment of positions.

Mean reversion in stock prices is a commonly accepted idea. Consistent with this notion, the Long Term Trend variable, which is modeled to account for this phenomenon, is statistically significant with a negative coefficient. This suggests that if the trend over the past year is positive (negative), then (holding all other variables constant at zero) the following day's return will be negative (positive).

The coefficient of the Volume variable is not statistically different from zero. Inclusion of the Volume variable thus confirms that the observed effects are not artifacts of changes in volume.

Lastly, strong support for the existence of resistance is exhibited in this data set. Indeed, the regression coefficient for the Resistance variable is both statistically significant and negative suggesting that a downward pressure is exerted on prices as they approach a recent high price.

**6. Conclusion**

This paper utilizes a methodology whereby the effects of price trend, resistance, volume changes, volatility and other variables on return are determined together. Without adjusting for valuation (thus using the raw data of individual S&P 100 stocks) most of these non-valuation effects would be masked by the noise inherent in the changes in valuation.

Another important aspect of this methodology that is neglected in many studies is that nonlinear effects are often invisible using linear methods, or they appear as small terms possibly with the opposite (or unexpected) sign.  In particular, raw data of major stocks usually shows a nearly negligible trend effect.  However, the combination of adjusting for changes in valuation and the examination of nonlinear terms establishes a nonlinear trend term with strong statistical significance and a magnitude that is comparable to the coefficient of valuation.  This suggests that for short term movements, the trend has an influence that is comparable to that due to changes in valuation.

The results show that the return is a cubic function of the trend.  In particular, the return is positively influenced by the trend, so long as the trend is one which is small as measured by standard deviation. Thus an uptrend that is observed sufficiently often that it is within 1.634 standard deviations tends to have a positive influence (see Figure 3), but beyond this point any increase in trend tends to reduce returns.  This effect can be attributed to underreaction, since it indicates investors are slow to react to a positive development.  For sufficiently large trends, there is a negative effect on return, providing evidence of overreaction.  Thus this methodology addresses these two key issues related to behavioral finance and provides answers for questions that critics have posed:  (1) Can one establish over- and underreaction?  (2) How does one distinguish between over- and underreaction?  The methodology indicates that traders are usually eager to observe an uptrend and to buy into it, but are also eager to take profits when the uptrend is significantly stronger than the typical level.

Another issue is whether this cubic is symmetric or asymmetric. Symmetry indicates a balance between buying on dips or taking profits, and is reflected in Figure 3. Asymmetry in the form of a negative root that is larger in magnitude than the positive root would indicate that traders are less eager to buy on dips than they are to take profits during that time period. This behavior could be an indication of a cautious or pessimistic outlook. The opposite situation would indicate a euphoric market. By examining different groups of stocks and eras, the shape of the cubic polynomial can be used to display motivations of traders and the temporal evolution of these motivations.

An important idea in behavioral finance involves anchoring, or using a particular number related to the trader's experience as a benchmark. Traders often utilize this concept in a practical context, and feel that a stock nearing its quarterly high, for example, will encounter resistance that will slow or terminate an uptrend. The methodology proposed and utilized in this paper shows that this is indeed the case.

This methodology facilitates the examination of a broad range of questions in a quantitative manner, and the independent variables are not limited to those considered above. Whether the hypothesized effect is fundamental or behavioral, it is possible to include it among the variables so long as it can be quantified. Hence, any theoretical model can be examined in terms of the predictions it renders, allowing researchers to accept, modify or reject facets of the model. A paradigm for behavioral finance thus involves an expanding and verifiable set of motivations that can be quantified and integrated into a model of market dynamics, much as physics has evolved with a growing list of experimentally established forms of energy.

An interesting application of this would be to examine different time periods e.g., the crisis period of 2008-2009, to determine how the cubic function formed by the regression coefficients changes. For example, are traders more eager to lock in profits, or are they more reluctant to buy on dips. Analysis of this type can provide a useful tool for trading and risk analysis.

The ideas can also be implemented in a different direction back-tested for trading purposes. At each time $t$, the data up until $t$ may be used in order to provide the optimal coefficients up to that time period. These coefficients can then be used to render a (out-of-sample) forecast for time $t+1$. The profit can then be determined based on these forecasts. Other tests such as the fraction of time in which the forecast has the right direction (i.e., positive or negative) can also be performed. Nevertheless, the purpose of this paper is not to develop a profitable trading strategy but rather to understand the motivations of traders for a particular class of assets during a particular time period.

Implementation of such a trading application would be improved by focusing on a specific group of stocks for which valuation can be estimated more precisely. For example, investors in companies manufacturing high-tech consumer goods react quickly to sales data and announcements of new products. Thus, quantifying and refining changes in valuation for a smaller set of stocks may lead to useful forecasting, in addition to an understanding of trader motivations.

Acknowledgments:  The authors are grateful to Professor Akin Sayrak for valuable comments.  A careful reading and numerous useful suggestions by three anonymous referees are also greatly appreciated.

**References**

Ang, A. & Bekaert, G. (2007), "Stock return predictability: Is it there?", *Review of Financial Studies*, Vol. 20 No. 3, 651-707.

Antoniou, A., Ergul, N., Holmes, P. & Priestley, R. (1997), "Technical analysis, trading volume and market efficiency: evidence from an emerging market", *Applied Financial Economics*, Vol. 7 No. 4, 361-365.

Bali, T. G., Demirtas, K. O. & Tehranian, H. (2008),"Aggregate earnings, firm-level earnings, and expected stock returns", *Journal of Financial and Quantitative Analysis*, Vol. 43 No. 3, 657-684.

Barberis, N., Shleifer, A. and Vishny, R. (1998),"A model of investor sentiment", *Journal of Financial Economics*, Vol. 49 No. 3, 307-343.

Black, F. (1986),"Noise", *Journal of Finance*, Vol. 41 No. 3, 529-543.

Bloomfield, R., Libby, R. & Nelson, M. W. (2000),"Underreactions, overreactions and moderated confidence", *Journal of Financial Markets*, Vol. 3 No. 2, 113-137.

Blume, L., Easley, D. & O'Hara, M. (1994), "Market statistics and technical analysis: The role of volume", *Journal of Finance*, Vol. 49 No. 1, 153-181.

Bodie, Z., Kane, A. & Marcus, A.J. (2008), *Investments*, 7[th] ed. McGraw-Hill Education, Boston.

Bremer, M., & Sweeney, R. J. (1991), "The reversal of large stock-price decreases", *The Journal of Finance*, Vol. 46 No. 2, 747-754.

Caginalp, G. & Ermentrout, B. (1990), "A kinetic thermodynamics approach to the psychology of fluctuations in financial markets", *Applied Mathematics Letters*, *3*, No. 4, 17-19.

Caginalp, G. & DeSantis, M. (2011),"Nonlinearity in the dynamics of financial markets", *Nonlinear Analysis: Real World Applications*, Vol. 12 No. 2, 1140-1151.

Caginalp, G. & DeSantis, M. (2011),"Stock price dynamics: nonlinear trend, volume, volatility, resistance and money supply", *Quantitative Finance*, Vol. 11 No. 6, 849-861.

Caginalp, G., DeSantis, M. & Sayrak, A. (2014),"The nonlinear price dynamics of US equity ETFs", *Journal of Econometrics*, Vol. 183, 193-201.

Campbell, J. & Thompson, S. (2008),"Predicting excess stock returns out of sample: Can anything beat the historical average?", *Review of Financial Studies*, Vol. 21 No. 4, 1509-1531.

Carhart, M. (1997), "On persistence in mutual fund performance", *Journal of Finance*, Vol. 52 No. 1, 57-82.

Chordia, T., Subrahmanyam, A. & Anshuman, V. (2001), "Trading activity and expected stock returns", *Journal of Financial Economics*, Vol. 59, 3-32.

Daniel, K., Hirshleifer, D. & Subrahmanyam, A. (1998), "Investor psychology and security market under- and overreactions", *Journal of Finance*, Vol. 53 No. 6, 1839-1885.

Davidson, R. & MacKinnon, J. G. (1993), *Estimation and inference in econometrics*, Oxford University Press, New York.

Fama, E. (1998),"Market efficiency, long-term returns, and behavioral finance", *Journal of Financial Economics*, Vol. 49 No. 3, 283-306.

Fama, E. & MacBeth, J. D. (1973),"Risk, return, and equilibrium: Empirical tests", *Journal of Political Economy*, Vol. 81, 607-636.

Gelman, A. (2008),"Scaling regression inputs by dividing by two standard deviations", *Statistics in Medicine*, Vol. 27 No. 15, 2865-2873.

George, T. J. & Hwang, C. Y. (2004),"The 52-week high and momentum investing", *Journal of Finance*, Vol. 59 No. 5, 2145-2176.

Goyal, A. (2012),"Empirical cross-sectional asset pricing: a survey", *Financial Markets and Portfolio Management*, Vol. 26 No. 1, 3-38.

Graham, B. & Dodd, D. L. (1934), *Security analysis*, McGraw Hill, New York.

Harris, M. & Raviv, A. (1993), "Differences of opinion make a horse race", *Review of Financial Studies*, Vol. 6 No. 3, 473-506.

Hong, H. & Stein, J. C. (1999), "A unified theory of underreaction, momentum trading, and overreaction in asset markets", *Journal of Finance*, Vol. 54 No. 6, 2143-2184.

Hsiao, C. (2003), *Analysis of Panel Data*, Cambridge University Press, New York.

Jegadeesh, N. & Titman, S. (1993),"Returns to buying winners and selling losers: Implications for stock market efficiency", *Journal of Finance*, Vol. 48 No. 1, 65-91.

Karpoff, J. (1987), "The relation between price changes and trading volume: A survey", *Journal of Financial and Quantitative Analysis*, Vol. 22 No. 1, 109-126.

Lee, C., Myers, J. & Swaminathan, B. (1999),"What is the Intrinsic Value of the Dow?", *Journal of Finance*, Vol. 54 No. 5, 1693-1741.

Leung, T. and Wang, Z., "Optimal risk-averse timing of an asset sale: Trending vs mean-reverting price dynamics. (2018).  SSRN: https://ssrn.com/abstract=2786276.

Luenberger, D.G. (1998), *Investment Science*, Oxford University Press, New York.

Madura, J. & Richie, N. (2004), "Overreaction of exchange-traded funds during the bubble of 1998-2002", *Journal of Behavioral Finance*, Vol. 5 No. 2, 91-104.

Menkhoff, L. (2010), "The use of technical analysis by fund managers: International evidence", *Journal of Banking & Finance*, Vol. 34, 2573-2586.

Moskowitz, T. & Grinblatt, M. (1999),"Do industries explain momentum?", *Journal of Finance*, Vol. 54 No. 4, 1249-1290.

Neely, C., Rapach, D., Tu, J. & Zhou, G. (2014), "Forecasting the equity risk premium: The role of technical indicators", *Management Science*, Vol. 60 No. 7, 1772-1791.

Petersen, M. (2009), "Estimating standard errors in finance panel data sets: Comparing approaches", *Review of Financial Studies*, Vol. 22 No. 1, 435-480.

Pinto, J.E., Robinson, T., and Stowe, J (2015), "Equity valuation: a survey of professional practice. SSRN: http://dx.doi.org/10.2139/ssrn.2657717

Poterba, J. M. & Summers, L. H. (1988),"Mean reversion in stock prices: Evidence and implications", *Journal of Financial Economics*, Vol. 22 No. 1, 27-59.

Sturm, R. R. (2003),"Investor confidence and returns following large one-day price changes", *Journal of Behavioral Finance*, Vol. 4 No. 4, 201-216.

Sturm, R. R. (2008), "The 52-week high strategy: momentum and overreaction in large firm stocks", *The Journal of Investing*, Vol. 17 No. 2, 55-67.

Sturm, R. R. (2013), "Market Efficiency and Technical Analysis: Can they Coexist?", *Research in Applied Economics* 5, No. 3, 1-16.

Subrahmanyam, A. (2010), "The cross-section of expected stock returns: What have we learnt from the past twenty-five years of research?", *European Financial Management*, Vol. 16, 27-42.

Thompson, S. (2011), "Simple formulas for standard errors that cluster by both firm and time", *Journal of Financial Economics*, Vol. 99, 1-10.

Vassalou, M. (2003), "News related to future GDP growth as a risk factor in equity returns", *Journal of Financial Economics*, Vol. 68, 47-73.

Wang, J. (1994), "A model of competitive stock trading volume", *Journal of Political Economy*, Vol. 102, 127-168.

**Appendix. Winsorization and Error Terms.**

The data discussed above involves winsorized data at the 1% level. In other words, data points that are outside of the (1%, 99%) interval are replaced by the last data points that are within that interval. The reasons for this are that we are interested in motivations of traders in normal times rather than in extreme times. In other words, it is possible that events that occur within 1% of the time would skew the regressions. Also, financial data are known to exhibit fat tails which could increase the problem. Nevertheless, the same regressions were performed for the non-winsorized data, and the results are quite similar, although the coefficients differ. The interpretation is that the regressions done with winsorized data (Table 2) reflect the trader motivations during trading days that are typical while the non-winsorized data (Table 4) take all points into account.

The choice of winsorization at 1% is typical of asset price change studies. It is also consistent with the distribution of the error terms in the regressions in that they are very close to the normal distribution between the 1% and 99% levels. In particular, an examination of the 257,683 error terms shows the following. For the normal distribution with mean 0 and standard deviation 0.01492, the values at the 1% and 99% levels are -0.0347 and +0.0347, respectively. For the error terms of the non-winsorized data, the corresponding values are -0.035 and +0.035. For the 10% and 90% levels, the values for the normal are ± 0.0191. The corresponding values for the error terms are ± 0.019. For the 0.1% and 99.9% the normal distribution values are ± 0.0461. The corresponding values for the error terms are ± 0.046. Finally, for the 0.01% level, the normal values are ± 0.0555 while the corresponding numbers for the error terms are ± 0.0550.

A plot of the cumulative distribution functions for the error terms and the normal distributions shows some deviation between the 99% and the 99.99% levels. The direction of the difference is toward fat tails as one might expect. The Ryan-Joiner normality test on the non-winsorized error terms (with N=257,365) shows a correlation of 0.888 between the error terms and the normal distribution.

Another way to look at the data is to note the value attained at the $k^{th}$ percentile, and to see how this differs from the theoretical (perfect Gaussian) $k^{th}$ percentile. For k=10% the observed data value is 0.019. For the Gaussian we have the corresponding value, V, is V(0.019) = 0.10143. The ratio of the two is 0.10143/0.1= 1. 0143, i.e., 1.43% difference from the ideal. At k=0.01%, the observed value is 0.055, while V(0.055) = 1. 1376×10$^{-4}$ differing from the ideal by 13.76%.

**Tables**

**Table 1.** Descriptive Statistics for Data Set

The data set consists of daily data for 85 firms corresponding to the time period December 28, 2004 through January 19, 2018.  It is a balanced panel data set with 3,289 records per firm.  Mean values are calculated on a per firm basis, and values reported in the table are calculated across all firms.

| | Mean | Min | First Quartile | Median | Third Quartile | Max |
|---|---|---|---|---|---|---|
| Mean Dollar Volume (Millions) | 577.13 | 77.98 | 288.42 | 383.09 | 637.28 | 5,056.11 |
| Mean Volume (# of shares in Millions) | 14.35 | 0.59 | 4.39 | 7.02 | 13.62 | 134.59 |
| Mean Market Capitalization (Millions) | 96,039.31 | 25,285.21 | 43,598.37 | 65,751.50 | 130,651.52 | 384,078.04 |

**Table 2.** Regression Results

Two-way fixed effects regressions with robust standard errors (in parentheses) clustered by firm are run on the balanced panel data set described in Section 4. The fixed effects are firm and time.

| | Model 1V | Model 1T | Model 2 | Model 3 | Model 4 |
|---|---|---|---|---|---|
| *Valuation* | 0.055 | | 0.395*** | 0.615*** | 0.560*** |
| | (0.087; 0.63) | | (0.114; 3.46) | (0.166; 3.7) | (0.165; 3.39) |
| $(Valuation)^2$ | | | | 0.112* | 0.093 |
| | | | | (0.065; 1.72) | (0.068; 1.36) |
| $(Valuation)^3$ | | | | -0.01 | -0.010 |
| | | | | (0.028; -0.36) | (0.028; -0.36) |
| *Price Trend* | | 0.180** | 0.584*** | 0.721*** | 0.654*** |
| | | (0.09;1.99) | (0.098;5.96) | (0.146; 4.94) | (0.145; 4.51) |
| $(Price\ Trend)^2$ | | | | 0.108 | 0.081 |
| | | | | (0.100; 1.08) | (0.103; 0.79) |
| $(Price\ Trend)^3$ | | | | -0.090*** | -0.090** |
| | | | | (0.035; 2.57) | (0.035; 2.57) |
| $Price\ Trend \times Valuation$ | | | | 0.103 | 0.091 |
| | | | | (0.157; 0.65) | (0.158; 0.58) |
| $Price\ Trend^2 \times Valuation$ | | | | 0.043 | 0.045 |
| | | | | (0.061; 0.7) | (0.062; 0.73) |
| $Price\ Trend \times Valuation^2$ | | | | 0.151*** | 0.153*** |
| | | | | (0.033; 4.58) | (0.033; 4.64) |
| *Short Term Volatility* | | | | | 0.101 |
| | | | | | (0.090; 1.1) |
| *Long Term Volatility* | | | | | 0.261*** |
| | | | | | (0.099; 2.63) |
| *Long Term Trend* | | | | | -0.290*** |
| | | | | | (0.042; -6.9) |
| *Volume* | | | | | 0.038 |
| | | | | | (0.049; 0.78) |
| *Resistance* | | | | | -1.31*** |
| | | | | | (0.487; -2.7) |
| *Intercept* | 0.0043 | 0.0043 | 0.0043 | 0.0043 | 0.0043 |
| R-Square | 0.4029 | 0.4029 | 0.4031 | 0.4037 | 0.4039 |
| No. Observations | 257,635 | 257,635 | 257,635 | 257,635 | 257,635 |
| No. Groups/Firms | 85 | 85 | 85 | 85 | 85 |
| No. Days (per Firm) | 3,031 | 3,031 | 3,031 | 3,031 | 3,031 |
| F Test for No Fixed Effects | 55.15*** | 55.13*** | 55.16*** | 55.15*** | 55.07*** |

Notes:

    a.   *, **, *** indicates significance at the 90%, 95%, and 99% level, respectively.

    b.   For each coefficient, the standard error (multiplied by 1,000) and t-value are denoted by ( ; ).

    c.   Coefficient values have been multiplied by 1,000 for exposition.

    d.   The intercept term, which is not corrected for heteroscedasticity, is calculated as the grand mean over all firms of the daily mean of the reported intercept and fixed-effects terms.

    e.   The R-Square measure developed by Theil (1961) is reported.

**Table 3.** List of firms

This table provides a listing of the firms used in this study.  The first 85 firms were used in the primary analyses (balanced panel of data).  The final 13 firms were excluded from the primary analysis due to lack of data, but are included for robustness.

| No. | Ticker | Name | Sector |
| --- | --- | --- | --- |
| 1 | AAPL | APPLE INC | Information Technology |
| 2 | ABT | ABBOTT LABORATORIES | Health Care |
| 3 | ACN | ACCENTURE PLC CLASS A | Information Technology |
| 4 | AGN | ALLERGAN | Health Care |
| 5 | AIG | AMERICAN INTERNATIONAL GROUP INC | Financials |
| 6 | ALL | ALLSTATE CORP | Financials |
| 7 | AMGN | AMGEN INC | Health Care |
| 8 | AXP | AMERICAN EXPRESS | Financials |
| 9 | BA | BOEING | Industrials |
| 10 | BAC | BANK OF AMERICA CORP | Financials |
| 11 | BIIB | BIOGEN INC | Health Care |
| 12 | BK | BANK OF NEW YORK MELLON CORP | Financials |
| 13 | BLK | BLACKROCK INC | Financials |
| 14 | BMY | BRISTOL MYERS SQUIBB | Health Care |
| 15 | BRKB | BERKSHIRE HATHAWAY INC CLASS B | Financials |
| 16 | C | CITIGROUP INC | Financials |
| 17 | CAT | CATERPILLAR INC | Industrials |

| 18 | CL | COLGATE-PALMOLIVE | Consumer Staples |
| 19 | CMCSA | COMCAST A CORP | Consumer Discretionary |
| 20 | COF | CAPITAL ONE FINANCIAL CORP | Financials |
| 21 | COP | CONOCOPHILLIPS | Energy |
| 22 | COST | COSTCO WHOLESALE CORP | Consumer Staples |
| 23 | CSCO | CISCO SYSTEMS INC | Information Technology |
| 24 | CVS | CVS HEALTH CORP | Consumer Staples |
| 25 | CVX | CHEVRON CORP | Energy |
| 26 | DHR | DANAHER CORP | Health Care |
| 27 | DIS | WALT DISNEY | Consumer Discretionary |
| 28 | DUK | DUKE ENERGY CORP | Utilities |
| 29 | EMR | EMERSON ELECTRIC | Industrials |
| 30 | EXC | EXELON CORP | Utilities |
| 31 | F | F MOTOR | Consumer Discretionary |
| 32 | FDX | FEDEX CORP | Industrials |
| 33 | FOX | TWENTY-FIRST CENTURY FOX INC CLASS | Consumer Discretionary |
| 34 | GD | GENERAL DYNAMICS CORP | Industrials |
| 35 | GE | GENERAL ELECTRIC | Industrials |
| 36 | GILD | GILEAD SCIENCES INC | Health Care |
| 37 | GOOGL | ALPHABET INC CLASS A | Information Technology |
| 38 | GS | GOLDMAN SACHS GROUP INC | Financials |
| 39 | HAL | HALLIBURTON | Energy |
| 40 | HD | HOME DEPOT INC | Consumer Discretionary |
| 41 | HON | HONEYWELL INTERNATIONAL INC | Industrials |
| 42 | IBM | INTERNATIONAL BUSINESS MACHINES CO | Information Technology |
| 43 | INTC | INTEL CORPORATION CORP | Information Technology |
| 44 | JNJ | JOHNSON & JOHNSON | Health Care |
| 45 | JPM | JPMORGAN CHASE & CO | Financials |
| 46 | KO | COCA-COLA | Consumer Staples |
| 47 | LLY | ELI LILLY | Health Care |
| 48 | LMT | LOCKHEED MARTIN CORP | Industrials |
| 49 | LOW | LOWES COMPANIES INC | Consumer Discretionary |
| 50 | MCD | MCDONALDS CORP | Consumer Discretionary |
| 51 | MDLZ | MONDELEZ INTERNATIONAL INC CLASS A | Consumer Staples |
| 52 | MDT | MEDTRONIC PLC | Health Care |
| 53 | MET | METLIFE INC | Financials |
| 54 | MMM | 3M | Industrials |
| 55 | MO | ALTRIA GROUP INC | Consumer Staples |
| 56 | MON | MONSANTO | Materials |
| 57 | MRK | MERCK & CO INC | Health Care |

| 58 | MS | MORGAN STANLEY | Financials |
| 59 | MSFT | MICROSOFT CORP | Information Technology |
| 60 | NEE | NEXTERA ENERGY INC | Utilities |
| 61 | NKE | NIKE INC CLASS B | Consumer Discretionary |
| 62 | ORCL | ORACLE CORP | Information Technology |
| 63 | OXY | OCCIDENTAL PETROLEUM CORP | Energy |
| 64 | PEP | PEPSICO INC | Consumer Staples |
| 65 | PFE | PFIZER INC | Health Care |
| 66 | PG | PROCTER & GAMBLE | Consumer Staples |
| 67 | QCOM | QUALCOMM INC | Information Technology |
| 68 | RTN | RAYTHEON | Industrials |
| 69 | SBUX | STARBUCKS CORP | Consumer Discretionary |
| 70 | SLB | SCHLUMBERGER NV | Energy |
| 71 | SO | SOUTHERN | Utilities |
| 72 | T | AT&T INC | Telecommunications |
| 73 | TGT | TARGET CORP | Consumer Discretionary |
| 74 | TWX | TIME WARNER INC | Consumer Discretionary |
| 75 | TXN | TEXAS INSTRUMENT INC | Information Technology |
| 76 | UNH | UNITEDHEALTH GROUP INC | Health Care |
| 77 | UNP | UNION PACIFIC CORP | Industrials |
| 78 | UPS | UNITED PARCEL SERVICE INC CLASS B | Industrials |
| 79 | USB | US BANCORP | Financials |
| 80 | UTX | UNITED TECHNOLOGIES CORP | Industrials |
| 81 | VZ | VERIZON COMMUNICATIONS INC | Telecommunications |
| 82 | WBA | WALGREEN BOOTS ALLIANCE INC | Consumer Staples |
| 83 | WFC | WELLS FARGO | Financials |
| 84 | WMT | WALMART STORES INC | Consumer Staples |
| 85 | XOM | EXXON MOBIL CORP | Energy |
| 86 | ABBV | ABBVIE INC | Health Care |
| 87 | AMZN | AMAZON COM INC | Consumer Discretionary |
| 88 | CELG | CELGENE CORP | Health Care |
| 89 | FB | FACEBOOK CLASS A INC | Information Technology |
| 90 | GM | GENERAL MOTORS | Consumer Discretionary |
| 91 | KHC | KRAFT HEINZ | Consumer Staples |
| 92 | KMI | KINDER MORGAN INC | Energy |
| 93 | MA | MASTERCARD INC CLASS A | Information Technology |
| 94 | PCLN | THE PRICELINE GROUP INC | Consumer Discretionary |
| 95 | PM | PHILIP MORRIS INTERNATIONAL INC | Consumer Staples |
| 96 | PYPL | PAYPAL HOLDINGS INC | Information Technology |

| 97 | SPG | SIMON PROPERTY GROUP REIT INC | Real Estate |
| 98 | V | VISA INC CLASS A | Information Technology |

**Table 4.** Regression Results for Non-winsorized data

Two-way fixed effects regressions for the non-winsorized data are run on the balanced panel data set described in Section 4. The fixed effects are firm and time. This is done for Model 4 which has the most complete set of variables.

|  | Model |
|---|---|
| $Valuation$ | 0.205 |
|  | (0.132; 1.56) |
| $(Valuation)^2$ | 0.232** |
|  | (0.11; 2.11) |
| $(Valuation)^3$ | 0.038 |
|  | (0.024; 1.61) |
| $Price\ Trend$ | 0.24 |
|  | (0.16; 1.5) |
| $(Price\ Trend)^2$ | 0.322*** |
|  | (0.079; 4.1) |
| $(Price\ Trend)^3$ | -0.00007 |
|  | (0.01; -0.01) |
| $Price\ Trend\ \times Valuation$ | 0.046 |
|  | (0.029; 1.57) |
| $Price\ Trend^{\ 2} \times Valuation$ | 0.101** |
|  | (0.04; 2.52) |
| $Price\ Trend\ \times Valuation^2$ | 0.513** |
|  | (0.201; 2.55) |
| $Short\ Term\ Volatility$ | 0.063 |
|  | (0.109; 0.58) |
| $Long\ Term\ Volatility$ | 0.211** |
|  | (0.098; 2.15) |
| $Long\ Term\ Trend$ | -0.35*** |
|  | (0.052; -6.87) |
| $Volume$ | 0.019 |
|  | (0.065; 0.29) |
| $Resistance$ | -1.33*** |
|  | (0.475; -2.8) |
| R-Square | 0.4039 |
| No. Observations | 257,635 |
| No. Groups/Firms | 85 |
| No. Days (per Firm) | 3,031 |
| F Test for No Fixed Effects | 55.07*** |

Notes:

a.  *, **, *** indicates significance at the 90%, 95%, and 99% level, respectively.

b.  For each coefficient, the standard error (multiplied by 1,000) and t-value are denoted by ( ; ).

c.  Coefficient values have been multiplied by 1,000 for exposition.

d.  The R-Square measure developed by Theil (1961) is reported.

**Figures**

**Figure 1.** Relationship between the following day's return and the Valuation and Price Trend variables. Using the (statistically significant) regression coefficients from Model 3, the return is plotted as a function of the Valuation and Trend.

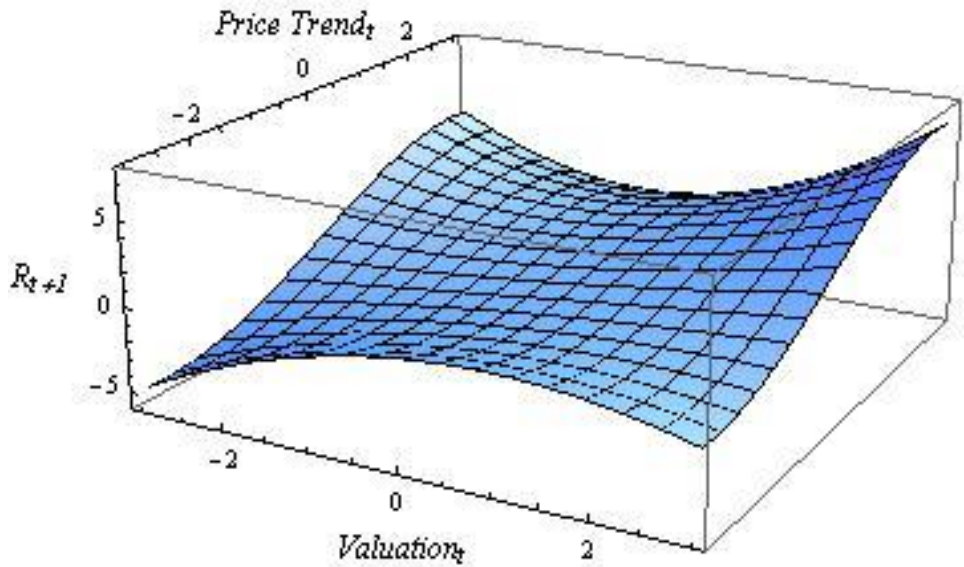

**Figure 2.** Plot of the following day's return versus valuation. Using the (statistically significant) regression coefficients from Model 3 with the change in Trend set to zero, the return is plotted as a function of the Valuation. The relationship between return and valuation is consistent with intuitive expectations of higher (lower) valuations corresponding to higher (lower) returns.

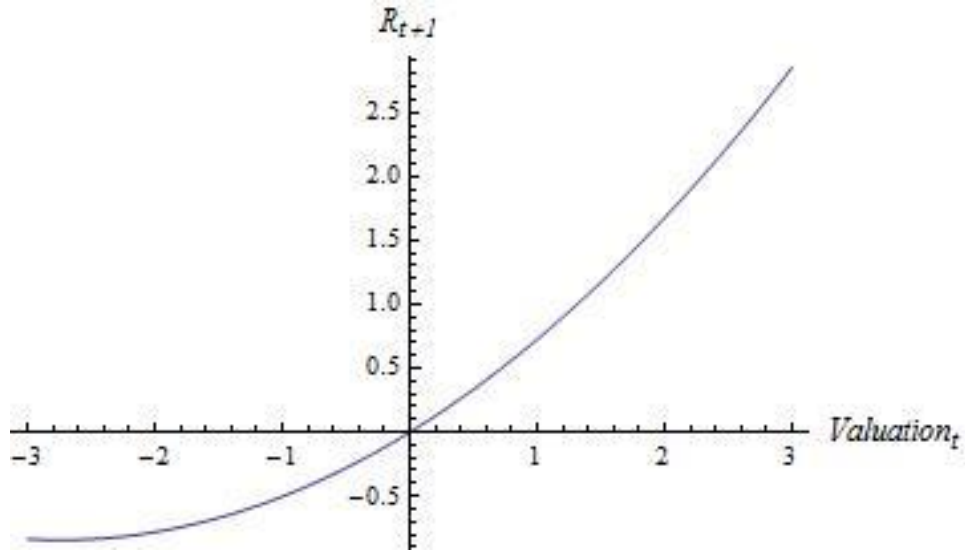

**Figure 3.** Plot of the following day's return versus trend in price. Using the (statistically significant) regression coefficients from Model 3 with the change in Valuation set to zero, the return is plotted as a function of the trend in price. Increasing price trends lead to greater returns for trends between -1.634 and 1.634 standard deviations. As the trend increases beyond 1.634 standard deviations, the return begins to diminish. Conversely, the return begins to increase as the trend decreases below -1.634 standard deviations.

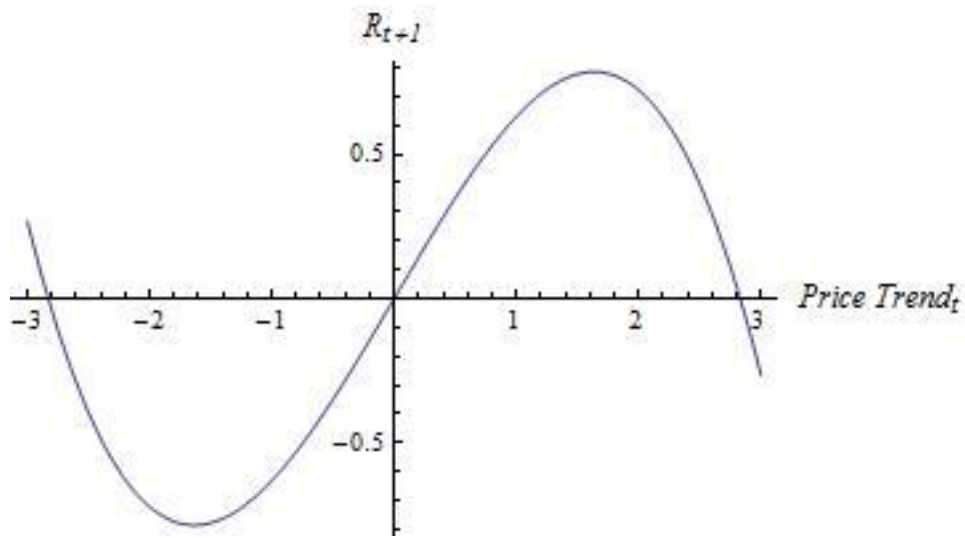